\documentclass[aps,pre,showkeys,groupedaddress,amsmath,amssymb,floatfix,twocolumn]{revtex4-1}

\usepackage{color} 
\usepackage{times}
\usepackage{todonotes}
\usepackage{verbatim}
\usepackage{bm}

\begin{document}


\title{Collective dynamics of random Janus oscillator networks}
\author{Thomas Peron$^{1,2}$}
\email{thomaskaue@gmail.com}
\author{Deniz Eroglu$^{3}$}
\email{deniz.eroglu@khas.edu.tr}
\author{Francisco A. Rodrigues$^{1}$}
\author{Yamir Moreno$^{2,4,5}$}
\affiliation{$^{1}$Institute of Mathematics and Computer Science, University of S\~ao Paulo,
S\~ao Carlos, S\~ao Paulo, Brazil}
\affiliation{$^{2}$Institute for Biocomputation and Physics of Complex Systems (BIFI), University of Zaragoza, Zaragoza 50018, Spain}
\affiliation{$^{3}$ Department of Bioinformatics and Genetics, Kadir Has University, 34083 Istanbul, Turkey}
\affiliation{$^{4}$Department of Theoretical Physics, University of Zaragoza, Zaragoza 50009, Spain}
\affiliation{$^{5}$ISI Foundation, Torino, Italy}

\begin{abstract}
Janus oscillators have been recently introduced as a remarkably simple phase oscillator model that exhibits non-trivial dynamical patterns -- such as chimeras, explosive transitions, and asymmetry-induced synchronization -- that once were only observed in specifically tailored models. Here we study ensembles of Janus oscillators coupled on large homogeneous and heterogeneous networks. By virtue of the Ott-Antonsen reduction scheme, we find that the rich dynamics of Janus oscillators persists in the thermodynamic limit of random regular, Erd\H{o}s-R\'enyi and scale-free random networks. We uncover for all these networks the coexistence between partially synchronized state and a multitude of 
states displaying global oscillations. Furthermore, abrupt transitions of the global and local order parameters are observed for all topologies considered. Interestingly, only for scale-free networks, it is found that states displaying global oscillations vanish in the thermodynamic limit.    
\end{abstract}

\maketitle 


Research on coupled oscillators in the past decade has been marked by the 
discovery of many intriguing patterns in the collective behavior of networks~\cite{arenas2008synchronization,rodrigues2016kuramoto}. Notable
examples of such patterns are chimeras~\cite{panaggio2015chimera}, states in which populations of synchronous are asynchronous oscillators coexist; explosive synchronization transitions~\cite{gomez2011explosive,rodrigues2016kuramoto}, which appear as a consequent of constraints in the natural frequency assignment; and asymmetry-induced synchronization~\cite{nishikawa2016symmetric,*zhang2017asymmetry}, a state in which synchrony is counter-intuitively favored by  oscillator heterogeneity.  In all these cases, phase oscillator models had to be specially designed so that those non-trivial states could be scrutinized. Very recently, however, Nicolaou et al.~\cite{nicolaou2019multifaceted} defined an oscillator model coined as Janus oscillators; the name is inspired in the homonym two-faced god of Roman mythology and reflects the two-dimensional character of an isolated oscillator -- each ``face'' of a Janus unit consists of a Kuramoto oscillator, whose natural frequency has the same absolute value but opposite sign to the frequency of its counter-face. When 
coupled on one-dimensional regular graphs, Janus oscillators have been found to exhibit a 
striking rich dynamical behavior that encompasses the co-occurrence of several dynamical patterns, 
in spite of the simplicity of the topology and the oscillator model itself~\cite{nicolaou2019multifaceted}.

    
The Janus model was introduced as a potential model for biological systems such as the Chlamydomonas cells with counterrotating flagella \cite{friedrich2012flagellar,wan2016coordinated}. It is thus important to understand the dynamics of a Janus system on related (realistic) topologies. Here we pose the question of whether the observed 1D rich collective dynamics exists on more complex networks of Janus oscillators. To address this issue, we employ the Ott-Antonsen (OA) ansatz~\cite{ott2008low} and obtain a reduced set of equations describing the system's evolution. From this reduced representation we find that, indeed, peculiar patterns of synchrony persist when Janus oscillators are placed on random regular, Erd\H{o}s-R\'enyi (ER) and scale-free (SF) random networks. We provide analytical and numerical evidence that the multitude of states in Janus dynamics is a consequence of the 
coexistence of infinite neutrally stable limit-cycle trajectories, which
we denominate ``breathing standing-waves''. Co-occurrence between 
classical partially synchronized and standing-waves are also reported. 
We further show that for high average degrees the collective states of ER networks are accurately described by the reduced system obtained
for random regular ones. Interestingly, we demonstrate that the coupling
range in which global oscillations are possible vanishes in the thermodynamic limit of SF networks.

We begin by defining the dynamics of $N$ Janus oscillators~\cite{nicolaou2019multifaceted} on heterogeneous networks as
\begin{equation}
\dot{\theta}_i = \omega_i + \sum_{j=1}^{2N} W_{ij}\sin(\theta_j - \theta_i),\; (i=1,...,2N)
\label{eq:JanusOnNetworks_Compact}
\end{equation}
where the $2N \times 2N$ matrix $\mathbf{W}$ is defined as
\begin{equation}
\mathbf{W}=\left[\begin{array}{cc}
0 & \beta\mathbf{\mathbf{I}+\sigma A}\\
\beta\mathbf{I}+\sigma\mathbf{A}^{{\rm T}} & 0
\end{array}\right].
\label{eq:matrix_W}
\end{equation}
Natural frequencies are assigned as $\omega_i = \omega_0  + \Delta/2$, for $i=1,...,N$; and $\omega_i = \omega_0 - \Delta/2$, for $i=N+1,...,2N$, where $\Delta$ is the frequency mismatch and $\omega_0$ is the average frequency, which we assume $\omega_0=0$. System (\ref{eq:JanusOnNetworks_Compact}) is analogous to a bipartite network or a multilayer network in which
oscillators belonging to the same group do not interact with one another, while
connections between groups are encoded in matrix $\mathbf{A}$. Notice
also that interactions between oscillators are weighted by the coupling strength 
$\sigma$, except for oscillators with indexes $(i,i+N)$, $i \in [1,N]$ -- these pairs of nodes interact with coupling strength $\beta$.

By defining the local order parameters
\begin{equation}
R_i = \sum_{j=1}^{2N} W_{ij}e^{{\rm i}\theta_j}, 
\label{eq:R_i}
\end{equation}
Eqs.~\ref{eq:JanusOnNetworks_Compact} are then decoupled as 
\begin{equation}
\dot{\theta}_i = \omega_i + \textrm{Im}[e^{-{\rm i} \theta_i}R_i].
\label{eq:decoupled_eqs}
\end{equation}


Following~\cite{barlev2011dynamics}, we consider 
an ensemble of systems defined by Eq.~\ref{eq:JanusOnNetworks_Compact} with fixed
coupling matrix $\mathbf{W}$. In this formulation, we describe the system (\ref{eq:JanusOnNetworks_Compact}) at a given time step $t$ by the joint probability density $\rho_{2N}(\bm{\theta},\bm{\omega},t)$, where $\bm{\theta}=(\theta_1,...,\theta_{2N})$ is the vector containing the 
phases at time $t$, and $\bm{\omega}=(\omega_1,...,\omega_{2N})$ is the time-independent vector with the natural frequencies of the individual oscillators.  
The evolution of the join probability $\rho_{2N}$ is then dictated by $\partial_t \rho_{2N} + \sum_{i=1}^{2N} \partial_{\theta_i}(\rho_{2N} \dot{\theta}_i)=0$, where $\dot{\theta}_i$
is given by Eq.~\ref{eq:decoupled_eqs}. Let us suppose that frequencies $\omega_j$ are distributed according to a generic function $g(\omega_j)$. Multiplying the continuity equation of $\rho_{2N}$ by $\Pi_{j \neq i} d\omega_j d\theta_j$ and integrating, 
we obtain the evolution equation for the marginal oscillator density $\rho_i(\theta_i,\omega_i,t)= \int \int \rho_{2N} \Pi_{j\neq i} d\omega_j \theta_j$; that is 
\begin{equation}
\frac{\partial \rho_i}{\partial t} + \frac{\partial }{\partial \theta_i}(\rho_i \dot{\theta}_i)=0.
\label{eq:one_oscillator_prob}
\end{equation}
By expanding $\rho_i$ in Fourier series and applying the OA ansatz to its coefficients we have~\cite{ott2008low,barlev2011dynamics}
\begin{equation}
\rho_{i}(\theta_{i},\omega_{i},t)=\frac{g(\omega_{i})}{2\pi}\left[1+\sum_{n=1}^{\infty}\hat{\alpha}_{i}^{n}(\omega_{i},t)e^{ \textrm{i} n \theta_{i}}+\rm{c.c}\right]
\label{eq:OA_ansatz}
\end{equation}
Inserting the previous equation into Eq.~\ref{eq:one_oscillator_prob}, we obtain the evolution for coefficients $\hat{\alpha}_i$:
\begin{equation}
\frac{d\hat{\alpha}_{i}}{dt}+{\rm i}\hat{\alpha}_{i}\omega_{i}+\frac{1}{2}\left[\hat{\alpha}_{i}^{2}R_{i}-R_{i}^{*}\right]=0\; (i=1,...2N),
\label{eq:alphas}
\end{equation}
where, in this ensemble approach, the coefficients $R_i$ are calculated as
\begin{equation}
R_i = \sum_{j=1}^{2N} W_{ij} \int_{-\infty}^{\infty} \int_0^{2\pi} \rho_j(\theta_j,\omega_j,t) e^{\textrm{i} \theta_j} d\theta_j d \omega_j.
\label{eq:Ri_ensemble}
\end{equation}
Inserting Eq.~\ref{eq:OA_ansatz} in the previous equation yields
\begin{equation}
R_i = \sum_{j=1}^{2N} W_{ij}\int_{-\infty}^{\infty} \hat{\alpha}_j^{*}(\omega_j,t)g(\omega_j) d\omega_j.
\label{eq:Ri_alpha}
\end{equation}

\begin{figure}[t!]
	\centering
	\includegraphics[width=0.75\columnwidth]{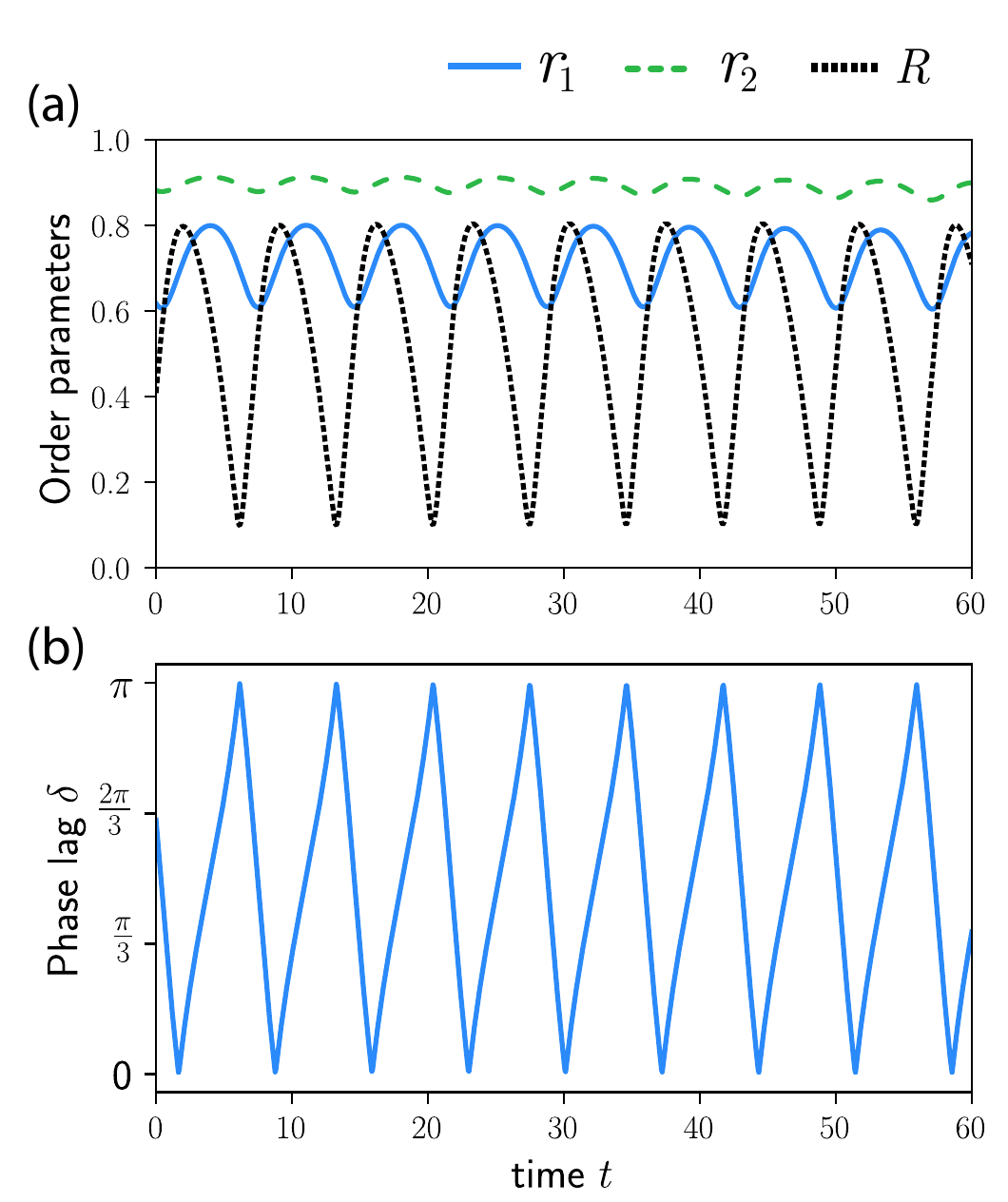}
	\caption{(Color online) Temporal evolution of (a) order parameters $r_{1,2}$ measuring the level of synchronization within subpopulations, and total order parameter $R = \frac{1}{2} \sqrt{r_1^2 + r_2^2 +2 r_1 r_2 \cos\delta}$; and (b) evolution of phase-lag $\delta$. Initial conditions: $r_1(0)=0.61$, $r_2(0)=0.83$, $\delta(0) = 2\pi/3$. Remaining parameters: $\beta = 0.0015$, $\Delta = 1$. Each subpopulation has $N = 2500$ oscillators.}%
	\label{fig:temporal}%
\end{figure}

\begin{figure}[t!]
	\centering
	\includegraphics[width=1.0\columnwidth]{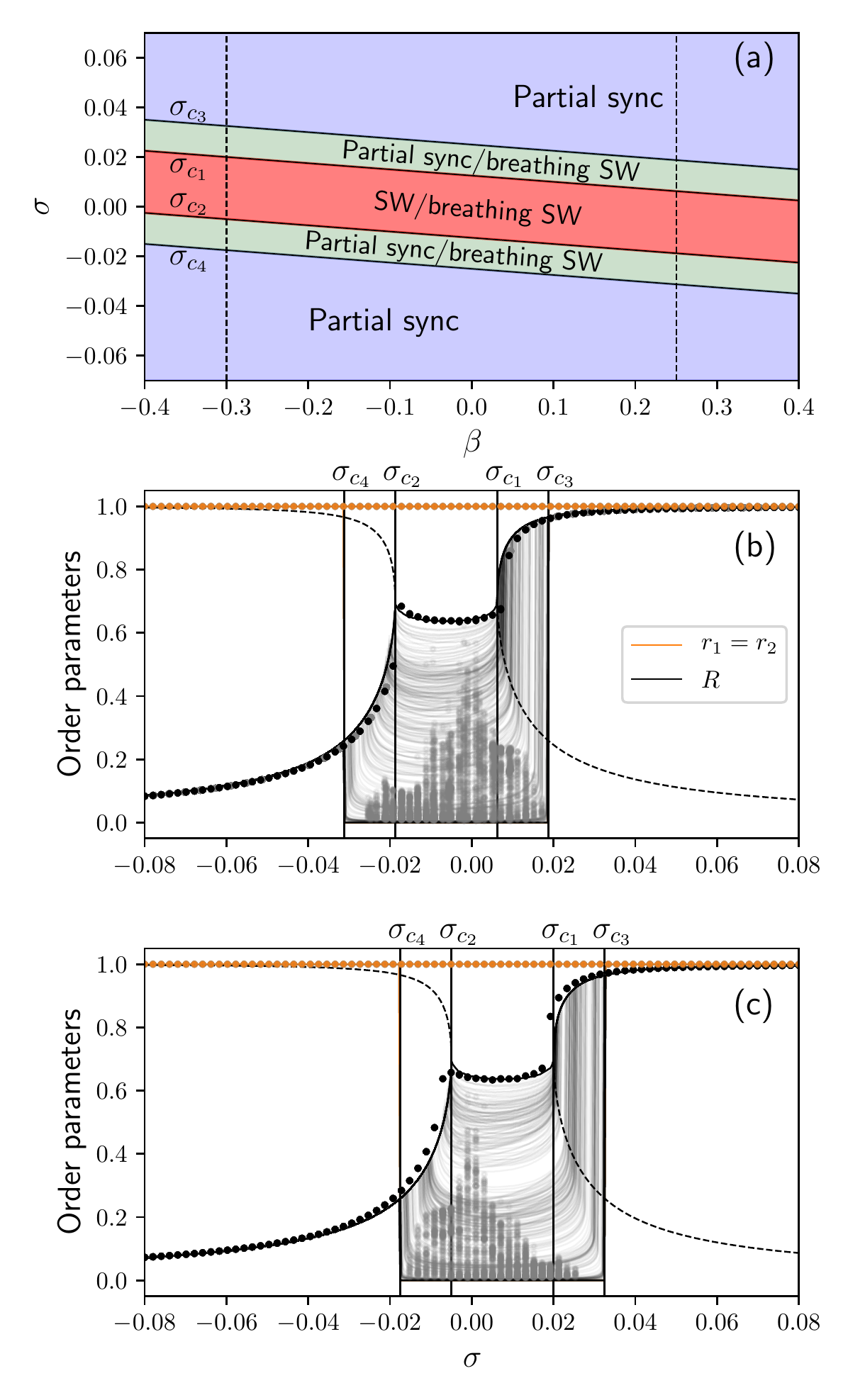}
	\caption{(Color online) (a) Stability diagram of system (\ref{eq:reduced_system_regular_networks}) for $k = 40$ and $\Delta = 1$. ``Partial sync'' refers to the state in which $r_{1,2} = 1$ and $\delta \neq 0$. SW denotes parameter regions for each $r_{1,2}=1$ and for which phase-lag $\delta$ rotates with a nonzero frequency. The regions where a multitude of solutions with $r_{1,2}<1$ and $\dot{\delta} \neq 0$ is found are labeled as ``breathing SW''. Order parameter curves for (b) $\beta = 0.25$ and (c) $\beta=-0.3$. Solid and dashed lines of $R$ for $\sigma \in (-\infty, \sigma_2] \cup [\sigma_1,\infty)$ 	are obtained with Eq.~\ref{eq:partial_synchronization_state_R}. The line for $r_1=r_2$ denotes the symmetric solution of Eq.~\ref{eq:reduced_system_regular_networks}. Dots are obtained by numerically evolving the original system (\ref{eq:JanusOnNetworks_Compact}) with $N=2500$ oscillators; each dot is an averaged over $t \in [250,500]$, with time step $dt = 0.05$.  Gray lines in the region $\sigma_{c_4}< \sigma < \sigma_{c_3}$ are generated numerically by evolving the reduced system in Eq.~\ref{eq:reduced_system_regular_networks} with random initial conditions. Gray dots depict the corresponding result yielded by the original system (Eq.~\ref{eq:JanusOnNetworks_Compact}). For the sake of clarity, we only show a small sample of the possible states attainable in the gray area.}%
	\label{fig:doubleRegular}%
\end{figure}


Let us now consider the case in which each subpopulation of the Janus coupling
scheme is a random regular network with degree $k$. More precisely, 
each oscillator of the subpopulation that rotates with frequency $\omega_1 = \Delta/2$ is
randomly connected to $k$ oscillators of subpopulation 2 ($\omega_2 = -\Delta/2$), and vice-versa. 
In this scenario, since oscillators within each group are identical -- i.e. $g(\omega_i)=\delta(\omega_i - \Delta/2)$, for $i=1,...,N$; and $g(\omega_i) = \delta(\omega_i + \Delta/2)$, for $i=N+1,...,2N$--, we assume 
the following solution for coefficients $\hat{\alpha}_i$:
\begin{equation}
\begin{aligned}
\hat{\alpha}_{1}=\hat{\alpha}_{2}=\cdots=\hat{\alpha}_{N}\equiv\alpha_{1}\\
\hat{\alpha}_{N+1}=\cdots=\hat{\alpha}_{2N}\equiv\alpha_{2}.
\end{aligned}
\label{eq:alpha_solution_regular}
\end{equation}
Hence, the local order parameters $R_i$ are reduced to:
\begin{equation}
R_{i}=\begin{cases}
\alpha_{2}^{*}(\beta+k\sigma) \equiv R_1 & \textrm{ if }i=1,...,N;\\
\alpha_{1}^{*}(\beta+k\sigma) \equiv R_2 & \textrm{ if }i=N+1,...,2N.
\end{cases}
\label{eq:R_double_regular}
\end{equation}
Inserting solutions (\ref{eq:alpha_solution_regular}) and (\ref{eq:R_double_regular}) 
in Eq.~\ref{eq:alphas}, we obtain the reduced set of equations
\begin{equation}
\dot{\alpha}_{1,2}+{\rm i}\omega\alpha_{1,2}+\frac{1}{2}\left[\alpha_{1,2}^{2}R_{1,2}-R_{1,2}^{*}\right]=0,
\label{eq:equation_alphas}
\end{equation}
which in terms of the coordinates $\alpha_{1,2} = r_{1,2} e^{i\psi_{1,2}}$ 
are written as
\begin{equation}
\begin{aligned}
\dot{r}_{1} & =  \frac{1 }{2}(\beta + k\sigma)r_{2}(1 - r_{1}^{2})\cos\delta\\
\dot{r}_{2} & =  \frac{1}{2}(\beta + k\sigma)r_{1}(1 - r_{2}^{2})\cos\delta\\
\dot{\delta} & =  -\Delta-\frac{1}{2}(\beta+k\sigma)\sin\delta\left[2r_{1}r_{2}+\frac{r_{1}}{r_{2}}+\frac{r_{2}}{r_{1}}\right],
\end{aligned}
\label{eq:reduced_system_regular_networks}
\end{equation}
where $\delta = \psi_1 - \psi_2$ is the phase-lag between subpopulations. Variables 
$r_1$ and $r_2$ turn out to be the order parameters measuring the level of synchronization
within each subpopulation in the Janus system; the traditional Kuramoto order parameter
evaluating the global synchrony is obtained through $R(t) = \frac{1}{2}|r_1 e^{i\psi_t(t)} + r_2 e^{i\psi_2(t)} |$. States that emerge from system (\ref{eq:reduced_system_regular_networks}) are summarized as: (1) a partially synchronized state in which $r_{1,2}=1$, while the subpopulations remain separated by a constant phase-lag $\delta$ (hence, $R<1$); (2) a standing-wave (SW) state, where the bulks of the two
fully synchronized populations ($r_{1,2}=1$) rotate in opposite directions yielding a incessantly rotating $\delta$; (3) a distinct form of SW emerges when $0< r_{1,2}< 1$: along
with the increment or decrease in phase-lag $\delta$, the order parameters $r_{1,2}$ exhibit a breathing
behavior, as depicted in the simulation shown Fig.~\ref{fig:temporal}. Henceforth we refer to this state as ``breathing SW''. As we shall see, the classical incoherent state remains unstable for all coupling values.

In order to uncover the conditions for the existence of the partially synchronized state, we set $r_1(t) = r_2(t) \equiv r (t)$ in Eqs.~\ref{eq:reduced_system_regular_networks}, leading to $\dot{r} = \frac{\gamma r}{2} (1 - r^2) \cos \delta$ and $\dot{\delta} = - \Delta - \gamma (1+r^2)\sin\delta$. Imposing $\dot{r} = 0$, we notice that $r=1$ is always a fixed point. Inserting
the latter solution in the equation for $\dot{\delta}=0$, we find
that $\sin \delta = -\Delta/2\gamma$. Thus, the partially synchronized regime remains stable 
when $-\Delta /2|\gamma| \leq 1$ is satisfied. In terms of coupling $\sigma$, we then write these critical conditions as
\begin{equation}
\sigma_{c_1} = -\frac{\Delta/2 + \beta}{k}\textrm{ and }\sigma_{c_2} = \frac{\Delta/2 - \beta}{k}.
\label{eq:c_partial_synchronized_state}
\end{equation}
Couplings $\sigma_{c_1}$ and $\sigma_{c_2}$ determine the coupling range where 
the partially synchronized state exists. The total order parameter $R$ is then given by 
\begin{equation}
R = \frac{1}{\sqrt{2}} \sqrt{1 \pm \sqrt{1 - \frac{\Delta^2}{4(\beta + k \sigma)^2}}},
\label{eq:partial_synchronization_state_R}
\end{equation}
where the  ``-'' branch is stable for $\sigma \leq \sigma_{c_1}$, whereas the ``+'' branch 
is stable in the region $\sigma \geq \sigma_{c_2}$. For $\sigma_{c_1}<\sigma < \sigma_{c_2}$, 
the limit cycle solution of $\dot{\delta}$ holds and SW states emerge with perfectly synchronized subpopulations ($r_{1,2}=1$).

A linear stability analysis of the incoherent state ($\alpha_{1,2}=0$) in Eqs.~\ref{eq:equation_alphas} reveals that the eigenvalues of the Jacobian matrix
become purely imaginary at $|\beta + k\sigma | = \Delta$. Therefore, limit cycle 
solutions arise for 
\begin{equation}
\sigma_{c_{3}}\equiv-\frac{(\Delta+\beta)}{k}<\sigma<\frac{\Delta-\beta}{k}\equiv\sigma_{c_{4}}.
\label{eq:conditions_limit_cycles}
\end{equation}


Figure~\ref{fig:doubleRegular}(a) outlines the critical conditions given by Eqs.~\ref{eq:c_partial_synchronized_state} and ~\ref{eq:conditions_limit_cycles} in the 
plane spanned by couplings $\beta$ and $\sigma$. As it can be seen, the partially synchronized 
state is favored by extreme values of both couplings, whereas states with oscillating 
synchrony appear for intermediate values in the parameter space. In Figs.~\ref{fig:doubleRegular} (b) and (c) we visualize the evolution of the local and global order parameters over 
two vertical sections of the diagram in (a), namely for $\beta = 0.25$ and $-0.3$, respectively.
Supposing we initiate the system with a negative $\sigma$ in the ``Partial sync'' region ($\sigma < \sigma_{c_4}$), the 
total order parameter $R$ collapses in the solution given by Eq.~\ref{eq:partial_synchronization_state_R}. As $\sigma$ is further increased, at $\sigma = \sigma_{c_2}$ the unstable and stable branches of $R$ merge via a saddle-node infinite-period bifurcation, whereby the limit cycle
solution of the SW arises (see Figs.~\ref{fig:doubleRegular} (b) and (c)). Upon further 
continuation of $\sigma$, a saddle-node appears again at $\sigma = \sigma_{c_1}$ and the system 
is brought back to the partially synchronized state.

Besides the branch of $R$ obtained under the symmetry condition $r_1 = r_2 = 1$, the numerical evolution of the reduced system predicts the existence of several other curves, which are upper bounded by $R$ of the $r_{1,2}=1$ solution for $\sigma \in [\sigma_{c_4}, \sigma_{c_3}]$. Insights about 
the nature of such states can be gained by investigating the stability of the SW state under perturbations 
transversal to the symmetric manifold $r_1 = r_2$~\citep{martens2009exact}. By defining the transversal 
and longitudinal coordinates $r_{\perp} = (r_1 - r_2)/2$ and $r_{\parallel} = (r_1 + r_2)/2$, we 
have $\dot{r}_{\perp}=\frac{\gamma}{2}(r_{\perp}^{2}-r_{\parallel}^{2}-1)r_{\perp}\cos\delta$, which 
in terms of variables $b_{\parallel} = r_{\parallel}^2$ and $b_{\perp} = r_{\perp}^2$ reads  
\begin{equation}
\dot{b}_{\perp}=\gamma b_{\perp}(b_{\perp}-b_{\parallel}-1)\cos\delta. 
\label{eq:b}
\end{equation}
Linearization at a point $(b_0,\delta_0)$ lying on a limit cycle solution of the manifold $r_1 = r_2$ 
yields the variation equation $\dot{\delta b}_{\perp} = \lambda_{\perp} \delta b_{\perp}$, where
\begin{equation}
\lambda_{\perp} = -(\beta + k\sigma) (1 + b_0) \cos \delta_0.
\label{eq:variational_equation}
\end{equation}
By averaging the previous equation over a period of oscillation and using the periodicity
of the limit cycle ($\langle (d/dt) \ln b_0\rangle =0$) we find $\langle\lambda_{\perp}\rangle = -2(\beta + k\sigma)\langle\cos\delta_0\rangle$. Numerical calculations with original system (\ref{eq:JanusOnNetworks_Compact}) for extensive parameter combinations show that $\langle \cos \delta_0 \rangle \approx 0$ (and, consequently, $\langle \lambda_{\perp} \rangle \approx 0$) in the region $\sigma \in [\sigma_{c_2},\sigma_{c_1}]$, suggesting then that the limit cycle solution of the SW is neutrally stable. Although our numerical estimate does not give us an exact proof of the stability of the limit cycle solution, it sheds light on the existence of the multitude of curves observed in Fig.~\ref{fig:doubleRegular}(b) and (c). Essentially, any perturbation  
of the SW state leads to a new limit cycle with $r_{1,2}<1$, since nearby trajectories are not attracted nor repelled, explaining the origin of the numerous solutions encountered 
in the coupling region encompassed by $\sigma_{c_4}$ and $\sigma_{c_3}$ in Fig.~\ref{fig:doubleRegular}. Notice also that the lower branches in the region $\sigma \in [\sigma_{c_4}, \sigma_{c_3}]$ do not correspond precisely to the classical incoherent state, but rather represent limit cycles solutions with small amplitudes $r_{1,2}$.
It is noteworthy mentioning also that although we have considered negative and 
positive couplings, we have not
observed in the populations of Janus oscillators states akin to traveling waves and $\pi$-states, which are collective phenomena that are characteristic of the interplay 
between attractive and repulsive interactions~\cite{hong2011kuramoto,sonnenschein2015collective}.  

\begin{figure}[t!]
	\centering
	\includegraphics[width=1.0\columnwidth]{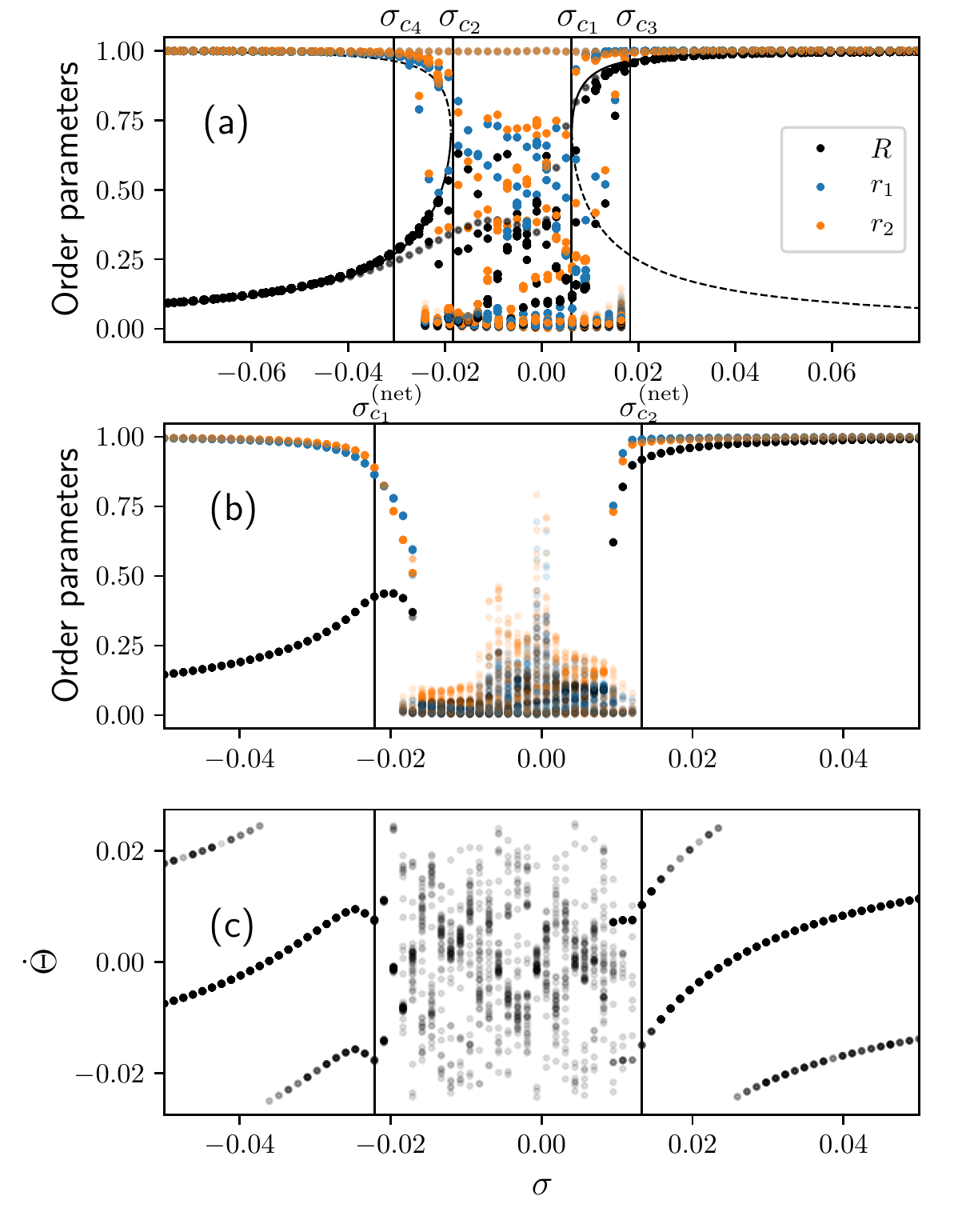}
	\caption{(Color online) (a) Evolution of order parameters $R$ and $r_{1,2}$ for ER networks with average-degree $\langle k \rangle = 40$. Dots are numerical experiments with $N = 2500$ oscillators. In order to highlight the dynamical similarity between random regular and dense ER networks, critical conditions and branches in this panel are obtained as in Fig.~\ref{fig:doubleRegular}. (b) Synchronization curves of an uncorrelated SF network with $p_k \sim k^{-\gamma}$, where $\gamma = 2.25$, $N = 10^{4}$ oscillators and minimum degree $k_{\min}=30$. Conditions $\sigma^{(\rm{net})}_{1,2}$ are given by Eq.~\ref{eq:conditions_limit_cycles_HMF}. (c) Evolution of mean-field  frequency $\dot{\Theta}$ associated to total order parameter $R(t)e^{i\Theta(t)} = (r_1 e^{i\psi_1} + r_2 e^{i\psi_2})/2$ for the same SF network of panel (b). Vertical lines in (c) are given by $\sigma^{(\rm{net})}_{1,2}$.}%
	\label{fig3}%
\end{figure}

The theory developed for random regular networks can also provide insights
on the dynamics of networks with mildly heterogeneous degree distributions. In Fig.~\ref{fig3}(a)
we superimpose numerical results for ER networks with the 
branches for the partially synchronized states (Eq.~\ref{eq:partial_synchronization_state_R}) 
and critical conditions given by Eq.~\ref{eq:c_partial_synchronized_state} and~\ref{eq:conditions_limit_cycles}. Interestingly, we see that the dependence of the order parameters
is reproduced with good precision with the expressions derived for simpler networks. Boundaries
enclosing the breathing SW states in the random regular network also delineate the region with global oscillations for the ER network. Notice also that $\sigma_{c_{1,2}}$ again mark the limits of the partially synchronize branch of $R$; however, no state analogous to the perfectly symmetric SW ($r_{1,2}=1$) is observed in $\sigma_{c_2} < \sigma < \sigma_{c_1}$ for ER networks.

Let us take a step further in the analysis of heterogeneous structures and consider general uncorrelated networks with degree distribution $p_k$. In this case, we assume that nodes with same degree $k$ admit the same solution, i.e., $\alpha_i = \alpha_k$, if $k_i=k$. Thus, Eqs.~\ref{eq:alphas} are reduced to
\begin{eqnarray*}
\dot{\alpha}_{k,1}+{\rm i}\omega\alpha_{k,1}+\frac{1}{2}\left[\alpha_{k,1}^{2}\left(\beta\alpha_{k,2}^{*}+\sigma\frac{k}{\langle k\rangle}\sum_{k^{\prime}}k^{\prime}p_{k^{\prime}}\alpha_{k^{\prime},2}^{*}\right)\right.\\
-\left.\left(\beta\alpha_{k,2}+\sigma\frac{k}{\langle k\rangle}\sum_{k^{\prime}}k^{\prime}p_{k^{\prime}}\alpha_{k^{\prime},2}\right)\right] & = & 0,
\label{eq:Janus_heterogeneous}
\end{eqnarray*}
where $\alpha_{k,1}$ describes the dynamics of oscillators with degree $k$ and frequency $\omega_i = \Delta/2$. Equations for coefficients $\alpha_{k,2}$
standing for the second face of Janus oscillators ($\omega_i = -\Delta/2$) 
are obtained accordingly. Linearizing the system around $\alpha_{k,1} = \delta \alpha_{k,1} \ll 1$ yields the following variational system 
\begin{equation}
\begin{aligned}
\dot{\delta\alpha}=-{\rm i}\omega\delta\bar{\alpha}+\frac{1}{2}\left[\beta+\sigma\frac{\langle k^{2}\rangle}{\langle k\rangle}\right]\delta\alpha\\
\dot{\delta\bar{\alpha}}=-{\rm i}\omega\delta\alpha-\frac{1}{2}\left[\beta+\sigma\frac{\langle k^{2}\rangle}{\langle k\rangle}\right]\delta\bar{\alpha},
\end{aligned}
\label{eq:perturbation_HMF}
\end{equation}
where $\delta \alpha$ is a small perturbation of the complex order parameter
$\alpha = \frac{1}{2\langle k \rangle} \sum_k k p_k (\alpha_{k,1} +\alpha_{k,2})$, and $\delta \bar{\alpha}$ is the analogous quantity for the parameter measuring the difference in the internal synchrony in Janus oscillators, i.e., $\bar{\alpha} = \frac{1}{2\langle k \rangle} \sum_k k p_k (\alpha_{k,1} - \alpha_{k,2})$. The eigenvalues of the Jacobian matrix of system (\ref{eq:perturbation_HMF}) become purely imaginary at $\Delta = |\beta + \sigma \langle k^2\rangle /\langle k \rangle|$. Therefore, we predict the appearance of states with global oscillations in the range
\begin{equation}
\sigma_{c_1}^{(\rm{net})}\equiv-(\Delta+\beta)\frac{\langle k \rangle}{\langle k^2 \rangle}<\sigma<\frac{\langle k \rangle}{\langle k^2 \rangle}(\Delta-\beta)\equiv \sigma_{c_2}^{\rm{(net})}.
\label{eq:conditions_limit_cycles_HMF}
\end{equation}
We check the predictions of the equation above in Fig.~\ref{fig3}(b) for 
SF networks with power-law exponent $\gamma = 2.25$. At first sight, it seems 
that the condition $\sigma_{c_1}^{(\rm{net})}$ provides an inaccurate estimation 
of the region where the order parameters $R$ and $r_{1,2}$ 
are expected to exhibit an erratic behavior, suggesting perhaps that finite-size
effects could be behind the deviation between $\sigma_{c_1}^{(\rm net)}$ and the 
point $\sigma \simeq 0.17$ at which the branch of $R$ collapses to values $R \approx 0$. However, Eq.~\ref{eq:conditions_limit_cycles_HMF} refers to the coupling range in which multiple
oscillating states are expected to emerge. Visualizing in Fig.~\ref{fig3}(c) the evolution
of the mean-field frequency $\dot{\Theta}$, we observe that $\sigma_{c_{1,2}}^{(\rm{net})}$ actually define
very accurately the boundaries of the states with multiple oscillating solutions ($\sigma_{c_1}^{\rm{(net)}}  < \sigma < \sigma_{c_2}^{\rm{(net)}}$). Given the dependence on $\langle k \rangle /\langle k^2\rangle$, one further envisions from Eq.~\ref{eq:conditions_limit_cycles_HMF} the absence of such oscillating states
in the thermodynamic limit for SF networks with $2 < \gamma \leq 3$, since $\sigma_{c_{1,2}}^{\rm{(net)}}$ are expected to vanish as $N \rightarrow \infty$, similarly to the classical coupling for the onset of synchronization in such structures~\cite{peron2019onset,rodrigues2016kuramoto}.



In conclusion, we have explored the collective dynamics of Janus oscillators
on large homogeneous and heterogeneous random networks. By employing the scheme 
provided by the OA ansatz, we obtained, for random regular networks, a reduced set of 
equations whereby critical points of the dynamics were revealed. We found that several collective behaviors coexist for intermediate coupling values, elucidating the findings in~\cite{nicolaou2019multifaceted}. Although initially obtained for homogeneous networks, we verified that the solutions of the reduced system fitted accurately numerical experiments for dense ER networks. By analyzing the 
stability of general uncorrelated networks, we further verified that the coupling range for which global oscillations are possible shrinks in the thermodynamic limit of SF networks. It is pertinent to mention that the accuracy of the OA ansatz in predicting the transition points is deteriorated for $\sigma$ and $\beta$ 
values beyond the region depicted in Fig.~\ref{fig:doubleRegular}. Deviations from the temporal 
signature yielded by the reduced system for solutions of $r_{1,2}$ were also  
observed in simulations for some couplings $(\beta,\sigma)$ in the breathing SW area. 

All in all, we provided the first theoretical and numerical analysis of ensembles of Janus oscillators on homogeneous and heterogeneous random networks. As such, our work raises further interesting questions about the study initiated by Nicolaou et al.~\cite{nicolaou2019multifaceted}. For instance, future investigations should target the dynamics on sparse and correlated networks -- situations in which the ensemble approach in~\cite{barlev2011dynamics}  
and mean-field techniques are inaccurate in predicting tipping points of the system -- as well as limitations of the OA manifold in capturing the Janus dynamics. 

TP acknowledges FAPESP (Grants No. 2016/23827-6 and 2018/15589-3). 
DE acknowledges Kadir Has University internal Scientific Research Grant (BAF). FAR acknowledges
CNPq (Grant No. 305940/2010-4) and FAPESP (Grants
No. 2016/25682-5 and grants 2013/07375-0) for the financial support given to this research. 
YM acknowledges partial support from the Government
of Aragon, Spain through grant E36-17R (FENOL), by
MINECO and FEDER funds (FIS2017-87519-P) and by
Intesa Sanpaolo Innovation Center. This research was
carried out using the computational resources of the Center for Mathematical Sciences Applied to Industry (CeMEAI) funded by FAPESP (grant 2013/07375-0). The
funders had no role in study design, data collection and
analysis, or preparation of the manuscript.


\bibliography{bibliography}

\end{document}